\documentclass[aps,prd, superscriptaddress,nofootinbib,showpacs]{revtex4}
\usepackage{amsmath,amssymb}

\usepackage{slashed}

\usepackage{graphicx}
\usepackage{color}

\newcommand{\p}{\partial}

\newcommand{\be}{\begin{equation}}
\newcommand{\ee}{\end{equation}}

\newcommand{\Lag}{\mathcal{L}}
\newcommand{\abcd}{{\alpha\beta\gamma\delta}}

\begin{document}

\title{On How Neutrino Protects the Axion}

\author{Gia Dvali}
\affiliation{Arnold Sommerfeld Center, Ludwig-Maximilians-University, Theresienstr. 37, 80333 M\"unchen, Germany}
\affiliation{Max-Planck-Institut f\"ur Physik. F\"ohringer Ring 6, 80805 M\"unchen, Germany}
\affiliation{Center for Cosmology and Particle Physics. Department of Physics, New York University. 4 Washington Place, New York, NY 10003, USA}
\author{Sarah Folkerts}
\email{sarah.folkerts@physik.lmu.de}
\affiliation{Arnold Sommerfeld Center, Ludwig-Maximilians-University, Theresienstr. 37, 80333 M\"unchen, Germany}
\author{Andre Franca}
\email{andre.franca@physik.lmu.de}
\affiliation{Arnold Sommerfeld Center, Ludwig-Maximilians-University, Theresienstr. 37, 80333 M\"unchen, Germany}

\begin{abstract}

$~~~$

We show how the neutrino can sacrifice itself to quantum gravity and save the axion solution to the strong-CP problem.  This mechanism puts an upper bound on the lightest neutrino mass.


\end{abstract}


\maketitle
\section{Introduction}

The celebrated solution to the strong CP problem is the Peccei-Quinn (PQ) solution \cite{2}. The essence of this solution is to promote the $\theta$-parameter
\be\label{eq:thetaQCD}
\Lag_\theta=\theta F\tilde{F}\;,
\ee 
into a dymanical field, a pseudo-Goldstone boson called axion \cite{Wilczek:1977pj,Weinberg:1977ma}. Here $F$ denotes the gluon field strength matrix and $\tilde{F}_{\mu\nu} = \epsilon_{\mu\nu\alpha\beta} F^{\alpha\beta}$ its dual.  The index structure in \eqref{eq:thetaQCD} is obvious and we do not display it explicitly.  The essence of this mechanism is as follows: The axion is sourced by $F\tilde{F}$ and its Lagrangian has the form 
\be\label{eq:axionLag}
\Lag_a=(\p_\mu a)^2+\frac{a}{f_a}F\tilde{F}\;,
\ee 
where $f_a$ is the axion decay constant\footnote{ Note that we will drop numerical factors in front of the kinetic terms in the following, which can easily be restored by taking into account the canonical normalization of the fields.}. Thus, $\theta$ effectively gets replaced by $\theta\to\frac{a}{f_a}$. The Lagrangian \eqref{eq:axionLag} guarantees that the minimum of the axion potential is necessarily at $\langle F\tilde{F}\rangle=0$ and therefore CP is unbroken. This also follows from the fact \cite{VW} that the
minimum of energy in QCD is at $\theta = 0$. 

Different axion models only differ by the underlying dynamics that yield the coupling $\frac{a}{f_a}F\tilde{F}$ in \eqref{eq:axionLag}. For example, in the original PQ formulation, the axion is the phase of a complex scalar field $\phi=\rho e^{i\frac{a}{f_a}}$ which spontaneously breaks an anomalous $U(1)_{PQ}$ symmetry
\be\label{eq:PQsymm}
\phi\rightarrow\phi e^{i\alpha}\;,
\ee 
through its vacuum expectation value (vev) $\langle|\phi|\rangle=f_a$. The axion $a$ then becomes a (pseudo) Goldstone boson. The coupling in \eqref{eq:axionLag} is generated through the chiral anomaly. 

However, for the solution of the strong CP problem the precise origin of the axion is unimportant. Any theory that delivers the coupling in \eqref{eq:axionLag} will solve the strong CP problem. In this paper, therefore, we shall treat the axion as an effective degree of freedom with Lagrangian \eqref{eq:axionLag} and shall not be interested in its microscopic origin. 

In this paper, we want to discuss a potential threat to the axion solution and a particular way to protect it. The threat appears in form of  an additional contribution  $V(a)$ to the axion potential 
\be\label{eq:threatpot}
\Lag_a=(\p_\mu a)^2+\frac{a}{f_a}F\tilde{F}-V(a)\;.
\ee
Such an additional potential can destabilize the solution. Indeed, from the axion equation of motion
\be\label{eq:axeom}
\Box a+\frac{d V}{d a}=\frac{1}{f_a}F\tilde{F}\;,
\ee
it is clear that the vacuum is no longer necessarily at $\langle F\tilde{F}\rangle=0$.

There are common beliefs that the source of such an unacceptably large $V(a)$ contribution arises from quantum gravity corrections, see e.g. \cite{6}. We must stress, however, that these arguments are mainly based on ``folks theorem'' about the non-existence of global symmetries in gravity.

The purpose of the present paper is not to contribute to such a debate. Instead, following \cite{3}, we would like to argue that for the axion such dangerous contributions can be limited to very well-defined effects. These can be consistently parametrized even in the absence of the full knowledge of gravity effects and a possible protection mechanism can be identified as well. This is due to the power of gauge invariance which allows to understand the axion solution in terms analogous to a Higgs (or St\"uckelberg) effect, but for a three-form (instead of one-form) gauge field. 

\section{Three-form Higgs effect}

We shall adopt the formulation of \cite{3} where it was shown that the axion as well as the massless quark solutions to the $\theta$ problem is fully equivalent to dynamically generating a mass gap for the Chern-Simons three-form $C$ of QCD 
\be\label{eq:Chernsimonsqcd}
C_{\alpha\beta\gamma} \, \equiv \,  {\rm Tr} \left (
A_{[\alpha}\partial_{\beta}A_{\gamma]} + {2 \over 3} A_{[\alpha}A_{\beta}A_{\gamma]}\,  \right ),
\ee
 where $A$ is the gluon field matrix. It is straightforward to check that under QCD gauge transformations $C$ shifts as 
 \be\label{eq:Cshift}
 C_{\alpha\beta\gamma}  \rightarrow  C_{\alpha\beta\gamma} \, + \, d_{[\alpha}\Omega_{\beta\gamma]}\;,
\ee
where $\Omega$ is a two-form. Notice that $F\tilde{F}=\mathrm{d}C\equiv E$. Thus $F  \tilde{F}$ is a field strength for $C$. It is important to notice that $C$ is not just a formal notation, but it acquires the meaning of a field in QCD.

In order to appreciate the role of $C$ for the strong CP problem let us recall the following crucial fact: Physics is $\theta$-dependent only if the topological susceptibility of the vacuum is nonzero 
\be\label{eq:FFcorr}
\langle F\tilde{F},F\tilde{F}\rangle_{q\to 0}=\mathrm{const}\neq 0\;,
\ee
where $q$ is the momentum.  In particular, this is the case in QCD with no massless quarks or axions.  This common knowledge will be our starting point.

Then, rewritten in terms of the Chern-Simons three-form $C$, \eqref{eq:FFcorr} implies 
\be\label{eq:CCcorr}
\langle C, C\rangle_{q\to 0}=\frac{1}{q^2}\;.
\ee
Thus, the $\theta$-dependence of physics is directly related to the mass of $C$.

Equation \eqref{eq:FFcorr} and \eqref{eq:CCcorr} illustrate two very important things:
Whenever physics is $\theta$-dependent, the three-form $C$ appears as a massless field in the effective theory, i.e., the correlator $\langle C, C\rangle$ has a massless pole at $q^2=0$. 
Correspondingly, for a massive $C$ the pole at $q^2=0$ is removed and physics becomes $\theta$-independent. 

The solution to the strong CP problem is therefore equivalent to the generation of a mass gap for $C$. This is exactly what the axion and the massless quark solutions accomplish. In both cases, the theory delivers a new pseudo-scalar degree of freedom that is eaten up by the three-form $C$ which subsequently becomes massive. In the case of the axion solution the eaten up pseudo-scalar is the axion and in the case of the massless quark solution it is the $\eta^\prime$ meson. 

Following \cite{5} we can formulate the mechanism for the generation of a mass gap purely in the language of topological quantities. The advantage of this language is that it is very general and treats both the axion and the massless quark solution on the same footing. 

Let us first start with a theory in which physics is $\theta$-dependent. Then, from \eqref{eq:CCcorr} it follows that $C$ is a massless field in a low energy theory. Thus its action has the form
\be\label{eq:masslessC}
\Lag=\frac{1}{\Lambda^4}E^2+\ldots\;,
\ee
where $\Lambda$ is the QCD scale and ``$\ldots$'' denotes terms that are higher order in $E$ and its derivatives.  In the absence of new   light states, these terms are obtained by integrating out the 
massive states of QCD and vanish in the $q=0$ limit. Hence, they are unimportant for our discussion as they cannot affect the existence of a massless pole for $C$. It is obvious that the above effective Lagrangian demonstrates the existence of such a pole, as 
it describes a theory of a massless three-form field. 
Thus, thanks to the power of the effective field theory and the topological description, we can make a definite statement about the existence of the mass-gap without any need of knowing the structure of higher order terms.  

Let us now add a sector to the theory which introduces an anomalous current $J_\mu$ with divergence
\be\label{eq:divJ}
\p_\mu J^\mu=F\tilde{F}=E\;.
\ee
We can then see that such a current automatically removes the massless pole and generates a mass gap for $C$. Notice that the anomaly generates a unique interaction between $J_\mu$ and $E$ \cite{5}
\be\label{eq:interJE}
E\frac{\p_\mu}{\Box}J^\mu\;,
\ee
so that \eqref{eq:masslessC} becomes 
\be\label{eq:lagJ}
\Lag=\frac{1}{\Lambda^4}E^2+\frac{E}{\Lambda^2}\frac{\p_\mu}{\Box}J^\mu\;.
\ee
(For simplicity, we set the anomaly coefficient equal to one). 
Performing the variation with respect to $C$, we get
\be\label{eq:eomwithJ}
\p_\nu E= -\Lambda^2\p_\nu\frac{\p_\mu}{\Box}J^\mu
\ee
and taking into account \eqref{eq:divJ} we obtain (up to an irrelevant constant)
\be\label{eq:higgsedE}
\Box E=-\Lambda^2 E\;.
\ee
It is clear that $E$ now propagates a massive field and the pole at $q^2=0$ is removed. Thus, physics becomes $\theta$-independent. The two known solutions to the strong CP problem act in precisely this way and only differ by the particular form of the current $J_\mu$. For the PQ axion, the current is given by
\be\label{eq:axionJ}
J_\mu=f_a\p_\mu a\;,
\ee
whereas for the massless quark case it is
\be\label{eq:massleJ}
J_\mu=\bar{Q}\gamma_\mu\gamma_5Q\;, 
\ee
where $Q$ is the massless quark. Notice,  in the latter case the role of the axion is played by the
$\eta'$-meson \cite{3}. 

\subsection{Goldstone-Higgs Interpretation of the Mass-Gap Generation} 

  Following \cite{3,5}, we would like to give a physical interpretation to the mass gap generation phenomenon that we 
  have just proven, in terms of the effect that is fully analogous to the ordinary Higgs effect  in spin-$1$ gauge theory. 
  There,  as we know, a massless spin-$1$ field that propagates two degrees of freedom eats-up  a pseudo-scalar  
with one propagating degree of freedom and forms a massive  spin-$1$ irreducible representation of the Poincare group, with  three propagating degrees of freedom. 

In \eqref{eq:lagJ}, we are facing a fully analogous phenomenon, but for of a three-form.   The would-be massless three-form field 
 that propagates zero-degrees of freedom  acquires one degree of freedom and becomes a massive pseudo-scalar.  
 The natural question to be asked is, where is this eaten-up pseudo-scalar coming from?  In the axion and the massless quark scenario the answer is very clear. The eaten-up scalar is a Goldstone boson of the spontaneously-broken Chiral symmetry corresponding to an anomalous current.  The corresponding Goldstone bosons are the axion and the $\eta'$-meson of QCD.  In both cases, the theory accommodates the existence of the Goldstone 
 bosons by consistency. This is ensured by the fact that symmetry corresponding to the anomalous current is also spontaneously  broken. For the axion this is accomplished by the vacuum expectation value of a Peccei-Quinn field $\phi$, whereas for the $\eta'$ meson it is achieved by the quark condensate of QCD.  
  This is very important for consistency since the new degree of freedom must come from somewhere.  The lesson that we are drawing from here is very important.  What we are finding is that whenever the anomalous current $J_{\mu}$ that  contributes to the generation of the mass-gap of three-form field  exists,  by consistency,  the quantity $\frac{\p_\mu}{\Box}J^\mu$ must act as an effective Goldstone 
 degree of freedom.  This statement is very general and depends neither on the microscopic nature of the three-form 
 nor on the origin of $J_{\mu}$.   The power of it is that it just relies on gauge invariance and the anomaly.  Any theory that violates this conditions cannot contribute to the generation of the mass gap.

\section{The role of gravity}
\label{sec:gravrole}

Now, we are fully equipped to monitor how gravity - or any other physics - could ruin the axion solution. The answer is simple: Gravity has to undo the generation of the mass gap for the three-form and recreate a massless pole at $q^2=0$. The only way to accomplish this is to create extra terms in equation \eqref{eq:divJ}. For example, this could be an additional potential term generated by gravity for the axion 
\be\label{eq:extrapot}
\Box a=\frac{1}{f_a}E-\frac{d V}{d a}\;.
\ee

In the PQ formulation of the problem such a potential term looks totally uncontrollable and could come from arbitrary $U(1)_{PQ}$ violating terms in the Lagrangian. This is not true in a dual formulation of the theory in which the scalar axion field $a$ is replaced by a two-form field $B_{\mu\nu}$. Due to the  gauge invariance of $B_{\mu\nu}$, the terms generated by gravity are extremely constrained\footnote{The UV-completion of the theory above the scale $f_a$ can, in principle, distinguish the two formulations.  We shall not address the issue of UV-completion in this paper. }.

To see this, we shall perform a duality transformation of the Lagrangian 
\be\label{eq:axLag}
\Lag=(\p_\mu a)^2+\frac{a}{f_a}E+\frac{1}{\Lambda^4}E^2\;.
\ee
by promoting $\p_\mu a\equiv R_\mu$ into a one-form and imposing the Bianchi identity  $\p_\alpha \epsilon^\abcd R_\delta =0$ via the Lagrange multiplier $B_{\mu\nu}$. Equation \eqref{eq:axLag} becomes 
\be\label{eq:oneform}
\Lag=R_\mu R^\mu + {1\over f_a} R_\alpha \epsilon^\abcd C_{\alpha\beta\gamma}\, +\, {1\over f_a} B_{\beta\gamma}\p_\alpha \epsilon^\abcd R_\delta +\frac{1}{\Lambda^4}E^2\;.
\ee
It follows from integrating out $R_\mu$ that the Lagrangian of the two-form field $B$ coupled to the three-form $C$ is
\be\label{eq:23form}
\Lag=\frac{1}{\Lambda^4}E^2+\frac{1}{f_a^2}(C-\mathrm{d}B)^2\;,
\ee
where we have dropped the indices and used a coordinate free representation of the $n$-forms. 

Notice that this system exhibits a gauge redundancy of the form
\begin{eqnarray}\label{eq:gaugetrafo}
&&C\to C+\mathrm{d}\Omega\;,\nonumber\\
&&B\to B+\Omega\;,
\end{eqnarray}
where $\Omega$ is an arbitrary two-form.

We would like to stress that the possibility of rewriting the QCD axion in a dual language of $B_{\mu\nu}$ and coupling it to a QCD Chern-Simons was already considered in \cite{6}, but unfortunately was abandoned, because the authors assumed that the duality between $a$ and $B_{\mu\nu}$ does not hold for a massive axion. This is not the case, since it is exclusively the coupling to $C$ that generates a mass for the axion in both formulations \footnote{Dual formulation of massive $B_{\mu\nu}$ in terms of massive 
three-form was  studied in \cite{quadr} up to quadratic order in fields.  Duality between the massive 
three-form and the massive axion for arbitrary form of the axion potential was proven in \cite{5}.}.

Equations \eqref{eq:23form} and \eqref{eq:gaugetrafo} exhibit the limitations for the possible gravitational damage in full glory. Since gravity has to respect the gauge redundancy \eqref{eq:gaugetrafo}, the only possibility to re-establish a massless pole is to provide another would-be massless three-form $C_G$ in such a way that a single axion is shared between $C$ and $C_G$ \cite{3}. 

In Einstein gravity the unique candidate for $C_G$ is the gravitational Chern-Simons term
\be\label{eq:gravcs}
C_G\equiv \Gamma \mathrm{d}\Gamma-\frac{3}{2} \Gamma\Gamma\Gamma\;,
\ee
with 
\be\label{eq:gravfieldstr}
\mathrm{d}C_G=R\tilde{R}\equiv E_G\;,
\ee
where $R$ is the Riemann tensor  and $\tilde{R}$ is its dual. For gravity thus to ruin the axion solution the following two conditions must be satisfied:
\begin{itemize}
\item The correlator 
\be\label{eq:RRcorr}
\langle R \tilde{R}, R\tilde{R}\rangle_{q\to 0}=\mathrm{const}\neq 0\;
\ee
in the absence of the axion.
\item The coupling $a R\tilde{R}$ or euqivalently $C_G\mathrm{d}B$ must be generated.
\end{itemize}

The above two quantities are parameterized by two  {\it a priori}-independent parameters.  The scale $\Lambda_G$ determines the strength of the correlator  
  $\langle R \tilde{R}, R\tilde{R}\rangle_{q\to 0}$, and will appear as an effective cutoff scale in a low energy theory 
  of the gravitational three-form $C_G$.  On the other hand, the strength of coupling to axion, which we shall denote by 
  $\alpha_G$, is determined by the 
  coefficient of the anomaly.   Whenever either of this two parameters is zero, gravity has no effect on the axion 
  mass.

The Lagrangian which renders the QCD $\theta$-term physical has the unique form
\be\label{eq:23forms}
\Lag=\frac{1}{\Lambda^4}E^2+\frac{1}{\Lambda_G^4}E_G^2+\frac{1}{f_a^2}(\alpha C+\alpha_G C_G-\mathrm{d}B)^2\;,
\ee
where as just explained $\alpha$ and $\alpha_G$ are two dimensionless parameters which determine the respective coupling strengths.  
 Normalizing the three-forms canonically, rescaling $B \rightarrow B\sqrt{(\alpha^2 \Lambda^4 + \alpha_G^2\Lambda_G^4)}$, and introducing the mixing angle, ${\rm cos}\beta \equiv {\alpha \Lambda^2 \over \sqrt{\alpha^2 \Lambda^4 + \alpha_G^2\Lambda_G^4}}$, we can rewrite, 
 \be\label{eq:23formscan}
\Lag=E^2+E_G^2+ m^2 ( C {\rm cos} \beta+ C_G{\rm sin}\beta-\mathrm{d}B)^2\;,
\ee
where $m^2 \equiv (\alpha^2 \Lambda^4 + \alpha_G^2\Lambda_G^4)/f_a^2$.  
We see that, due to mixing,  only one combination of three-forms is becoming massive, whereas 
the orthogonal one, $C {\rm sin} \beta - C_G{\rm cos}\beta$, is massless. The natural value of the QCD 
$\theta$-term is measured by the relative weight of the QCD three-form in this combination, 
 \be\label{eq:effectivetheta}
\theta_{QCD} \, = \,  {\rm sin} \beta \, \equiv \, {\alpha_G \Lambda_G^2 \over \sqrt{ \alpha^2 \Lambda^4 + \alpha_G^2\Lambda_G^4}} \,.
\ee    

We shall assume that this value is unacceptably large and look for possible solutions to comply with observations. An obvious way out is suggested by \eqref{eq:23forms}. We need to keep the gravitational three-form $C_G$ ``busy'' by coupling it to another two-form $B_{\mu\nu}^\prime$ 
\be\label{eq:22forms}
\Lag=\frac{1}{\Lambda^4}E^2+\frac{1}{\Lambda_G^4}E_G^2+\frac{1}{f_a^2}(\alpha C+\alpha_G C_G+\mathrm{d}B)^2+\frac{1}{f_a^{'2}}(C_G+\mathrm{d}B^\prime)^2\;.
\ee
where we have again restored the non-canonical normalization.  
In the usual terms this means that we need a new chiral current $J^\prime_\mu$ which is anomalous with respect to gravity. 

We would like to argue that the Standard Model contains a natural candidate for such a current in form of the neutrino lepton number. If this protection mechanism is to work, one obtains an upper bound on the lightest neutrino mass.

\section{Neutrino protection for the axion}
\label{sec:neutprot}

We will consider the Standard Model coupled to gravity. Our only assumption is that, in the absence of other chiral currents, gravity disturbs the axion solution. As discussed above, this implies that in the absence of the axion both correlators \eqref{eq:CCcorr} and \eqref{eq:RRcorr}  are nonzero. Equivalently, in the absence of any anomalous currents physics would depend on two $\theta$-angles $\theta_{QCD}$ and $\theta_{G}$. In this case, the low energy theory contains two massless three-forms, $C$ and $C_G$,
\be\label{eq:2massless3forms}
\Lag=\frac{1}{\Lambda^4}E^2+\frac{1}{\Lambda_G^4}E_G^2\;.
\ee
 Notice, below we shall work with a non-canonical normalization of the three-forms, but shall also   translate the results for the canonical normalization case. 


If gravity is to interfere with the axion solution, another necessary condition is that the axial current $J_\mu= f_a\p_\mu a$ is anomalous with respect to both QCD and gravity, i.e.
\be\label{eq:curranom}
\p_\mu J^\mu = \alpha E+\alpha_G E_G\;,
\ee
where $\alpha_G$ is associated with the gravitational breaking of the chiral symmetry.

 Since we are working at the level of effective low energy theory, we have no precise information about  $\alpha_G$, which depends on the UV structure of the theory, such as, e.g.,  the anomaly coefficients of the integrated-out heavy fermions. Therefore, we shall simply parametrize our ignorance and treat it as a free parameter. Of course, as explained above, our efforts for protecting the axion are required only if $\alpha_G$ is non-zero, which we shall assume in the following.  
 It is apparent from \eqref{eq:curranom} that for $\alpha_G\neq 0$ only one combination, namely 
\be\label{eq:higgsedcomb}
\alpha  E+\alpha_G E_G\;,
\ee
%
%
becomes massive. 
Indeed, the anomaly-generated effective Lagrangian 
\be\label{eq:efflag1}
\Lag=\frac{1}{\Lambda^4}E^2+\frac{1}{\Lambda_G^4}E_G^2+\frac{\alpha}{f_a^2}E\frac{\p_\mu}{\Box}J^\mu+\frac{\alpha_G}{f_a^2}E_G\frac{\p_\mu}{\Box}J^\mu 
\ee
implies the equations of motion 
\begin{eqnarray}\label{eq:eomforCGneut}
\partial_{\nu}  \frac{E}{\Lambda^4} &=& -\partial_{\nu} \left( \frac{\alpha}{f_a^2}\frac{\p_\mu}{\Box}J^\mu \right)\nonumber\\
\partial_{\nu} \frac{E_G}{\Lambda_G^4} &=& -\partial_{\nu} \left( \frac{\alpha_G}{f_a^2}\frac{\p_\mu}{\Box}J^\mu \right)\;,
\end{eqnarray}
which after taking into account  (\ref{eq:curranom}) can be rewritten as,  
\begin{eqnarray}\label{eq:equationsAAA}
\square( \alpha E + \alpha_G E_G) &=& -m^2  ( \alpha E + \alpha_G E_G) \nonumber\\
\square \left( \alpha_G {E \over \Lambda^4}  - \alpha {E_G \over \Lambda_G^4} \right ) &=& 0 \, .  
\end{eqnarray}
Where $m^2$ is defined below equation 
(\ref{eq:23formscan}). It is obvious that canonically normalized massive and massless 
combinations are given by the orthogonal superpositions determined by the same angle $\beta$ as 
defined before the equation (\ref{eq:23formscan}).  In particular, the massless pole at $q^2=0$ persists for $\alpha_G {E \over \Lambda^4}  - \alpha {E_G \over \Lambda_G^4}$, which in the language of canonically-normalized three-forms is
\be\label{eq:coulombcomb}
 {\rm sin}\beta  E- {\rm cos}\beta E_G\;.
\ee
 The non-fine-tuned value for $\theta_{QCD}$ is then given by  (\ref{eq:effectivetheta}), which 
 can be unacceptably large.  
 
However, the Standard Model contains other anomalous currents, such as the neutrino lepton number current. The existence of this current and the fact that it is anomalous under gravity changes the story 
and offers a protection mechanism. 

For definiteness, we consider a single massless neutrino species of left chirality $\nu_L$. The chiral symmetry 
\be\label{eq:chirsymm}
\nu_L\to e^{i\phi}\nu_L
\ee
is anomalous with respect to gravity and the corresponding current
\be\label{eq:neutcurr}
J^L_\mu=\bar{\nu}_L\gamma_\mu\nu_L\;,
\ee
has an anomalous divergence \cite{luis}
\be\label{eq:divneutJ}
\p_\mu J^{L\mu}= R\tilde{R}= E_G\;, 
\ee
 where we have set the known  anomaly coefficient to one. 
It is now obvious that the massless pole is no longer there for $E$ and $E_G$. Indeed from the effective Lagrangian 
\be\label{eq:efflagcur}
\Lag=\frac{1}{\Lambda^4}E^2+\frac{1}{\Lambda_G^4}E_G^2+\frac{\alpha}{f_a^2}E\frac{\p_\mu}{\Box}J^\mu+\frac{\alpha_G}{f_a^2}E_G\frac{\p_\mu}{\Box}J^\mu+\frac{1}{f_{\nu}^2}E_G\frac{\p_\mu}{\Box}J^{L\mu}
\ee
 (where $f_{\nu} \sim \Lambda_G$  is a scale that sets the strength of anomaly-induced coupling to neutrino current) one can obtain the equations of motion for $C$ and $C_G$ which together with \eqref{eq:divJ} and \eqref{eq:divneutJ} show that there are no massless modes in $E$ and $E_G$. 
 Thus, massless neutrino protects the axion solution to the strong-CP problem. In the next section we shall take into the account a possible effect of the small neutrino mass.


\subsection{Neutrino masses}
\label{sec:neutrinomasses}

Observations of neutrino oscillations \cite{nuosc} have established that there is an upper bound on neutrino masses $\sum m_\nu\lesssim 0.3 \mathrm{eV}$ \cite{neutrinomassbound}, where the sum is over all neutrino flavors. Therefore, neutrinos are interesting candidates for protecting the axion. 
Below we shall explore such a scenario. 

We therefore parametrize the mass of the lightest neutrino by $m_\nu$. In case $m_\nu$ is non-zero, the neutrino lepton number is explicitly broken. This introduces an additional factor to the divergence of the current \eqref{eq:neutcurr}
\be\label{eq:neutcurrentmass}
\p_\mu J^\mu_{\nu_L}= R\tilde{R}\, + \, m_\nu\bar{\nu}\gamma^5\nu\;.
\ee
 We shall not be concerned with the microscopic origin of this mass and shall treat it as a parameter.  
As we have shown before,  the generation of the mass gap for $C_G$ automatically implies that,  in full analogy with the \eqref{eq:axionJ} and \eqref{eq:massleJ}, this current must be 
identified with a pseudo-scalar degree of freedom. We shall denote it by  $\eta_{\nu}$ in analogy with the $\eta'$ meson of QCD.

  Notice, that we are not making any extra assumption. 
The necessity of a physical $\eta_{\nu}$ degree of freedom follows from the matching of high-energy 
and low-energy  theories \cite{3,5}.  From the high energy point of view, the axion is protected because the 
would-be gravitational  $\theta$-term is rendered unphysical by a chiral neutrino rotation.  In order to match this effect in low energy description, the correlator 
\eqref{eq:RRcorr} must be screened.  By gauge symmetry, this is only possible if there is a 
corresponding  Goldstone-type degree of freedom that plays the role of the St\"uckelberg field for the gravitational 
three-form (\ref{eq:gravcs}).  
 In other words, the same physics that provides the correlator \eqref{eq:RRcorr}, by consistency, must also provide the physical degree of freedom, $\eta_{\nu}$, necessary  for generation of the mass-gap in the presence of anomaly. 
 
 We can think  of $\eta_{\nu}$ as of a pseudo-Goldstone boson of the spontaneously-broken lepton-number symmetry  \eqref{eq:chirsymm} by non-perturbative gravity. 
   In a sense, $\eta_{\nu}$ can be thought of as a low-energy limit of the neutrino bilinear operator, 
\be\label{eq:etanudef}
\eta_\nu \rightarrow \frac{1}{\Lambda_G^2}\bar{\nu}\gamma^5\nu \quad \mathrm{and} \quad J^\mu_{\nu_L}\rightarrow \Lambda_G\p^\mu\eta_\nu\;, 
\ee
 in a way similar to relation of $\eta'$ of QCD in terms of a quark bilinear operator. With this connection, the effect 
 of a small neutrino mass on $\eta_{\nu}$  is similar to the effect of a small quark mass on $\eta'$. Namely, 
 to the leading order in $m_{\nu} \over \Lambda_G$  such a deformation of the theory should result into a small 
 explicit mass of $\eta_{\nu}$ in effective low energy Lagrangian.  Because physics must be periodic in the Goldstone field,  this mass term should be thought of as the leading order term in an expansion of the periodic function. The higher order terms in this expansion cannot affect the mechanism of mass-generation and are unimportant for the present discussion. 
Thus, the  dynamics of the theory is now governed by the Lagrangian \eqref{eq:efflagcur} with an additional mass term $m_\nu \Lambda_G \eta_\nu^2$ appearing from the explicit symmetry breaking by the neutrino mass. Replacing the currents with their corresponding pseudo-Goldstone bosons \eqref{eq:axionJ} and \eqref{eq:etanudef} yields
\begin{eqnarray}\label{eq:ac22formsmass}
\Lag &=& \frac{1}{\Lambda^4}E^2 +\frac{1}{\Lambda_G^4} E_G^2 -\frac{a}{f_a}E- \alpha_G\frac{a }{f_a}E_G-\frac{\eta_\nu}{\Lambda_G}E_G \nonumber\\
&&+ \p_\mu a\p^\mu a+\p_\mu \eta_\nu\p^\mu\eta_\nu- m_\nu \Lambda_G\eta_\nu^2\;.
\end{eqnarray}
 Here we have absorbed $\alpha$ into the definition of $f_a$ and have set the decay 
constant of $\eta_{\nu}$ to be equal to $\Lambda_G$ \footnote{In the absence of other scales in the problem, this is the only natural possibility,  in full analogy  to the decay constant of 
$\eta'$ meson being set by the QCD scale.  The possible difference between  $f_{\nu}$ and 
$\Lambda_G$ can easily be taken into the account and changes nothing in our analysis. }.  
  Ignoring numerical factors, the equations of motion for $C$ and $G$ are 
\begin{eqnarray}\label{eq:eomforCGneut}
\mathrm{d}\left( E -\Lambda^4\frac{a}{f_a}\right)&=&0\nonumber\\
\mathrm{d}\left( E_G -\alpha_G \Lambda_G^4\frac{a}{f_a}-\Lambda_G^3\eta_\nu\right)&=&0\;,
\end{eqnarray}
and the ones for $a$ and $\eta_\nu$ read 
\begin{eqnarray}\label{eq:eomforaetaneut}
f_a\Box a&=&-\alpha_G E_G-E\nonumber\\
(\Box+m_\nu \Lambda_G)\eta_\nu&=&-\frac{E_G}{\Lambda_G}\;.
\end{eqnarray}
 It is already clear from the last two equations that small enough neutrino mass will continue 
to keep $\theta$-term of QCD under control. Indeed, these two equations imply that in the vacuum 
(that is, for $\eta_{\nu} = a=$ constant) the value of the QCD electric four-form is
$E = m_{\nu} \alpha_G\Lambda_G^2 \eta_{\nu}$. Since, the vacuum expectation value of 
$\eta_{\nu}$ cannot exceed its decay constant, $\Lambda_G$, the corresponding maximal possible value of the  $\theta_{QCD}$ is 
\be\label{eq:thetamax}
\theta_{max}=  {E_{max} \over \Lambda^4} = {m_{\nu} \alpha_G\Lambda_G^3 \over \Lambda^4} \,, 
\ee
which vanishes for $m_{\nu} \rightarrow 0$.

Indeed, the vacuum solutions of (\ref{eq:eomforCGneut})  and (\ref{eq:eomforaetaneut})
 are given by the following expressions 
\begin{eqnarray}
\label{eq:exprforEEG}
E=\alpha_G\Lambda^4\frac{m_\nu \Lambda_G^4(\alpha_G \beta_2-\beta_1)}{\Lambda_G^4m_\nu \alpha_G^2+\Lambda^4(\Lambda_G+m_\nu)}\nonumber\\
E_G=\Lambda_G^4\frac{m_\nu\Lambda^4(\beta_1 - \alpha_G \beta_2)}{\Lambda_G^4m_\nu \alpha_G^2 +\Lambda^4(\Lambda_G+m_\nu)}\;,
\end{eqnarray}
where $\beta_1$ and $\beta_2$ are dimensionless integration constants of \eqref{eq:eomforCGneut}, 
i.e. $E= \Lambda^4\frac{a}{f_a}+\Lambda^4\beta_2$ and $E_G=\Lambda_G^4\alpha_G\frac{a}{f_a}+\Lambda_G^3\eta_\nu+\beta_1 \Lambda_G^4$.  Since the maximal values of the two electric 
fields are bounded by the scales $\Lambda^4$ and $\Lambda_G^4$,  the maximal 
values of the crresponding integration constants can be order one. 

Let us parametrize this result by the ratio of the neutrino mass to the gravitational scale by defining $\epsilon \equiv\frac{m_\nu}{m_{\nu}+\Lambda_G}$. 
 The maximal value for the quantity $\alpha_G \beta_2-\beta_1$ is either order one or  $\alpha_G$ depending whether $\alpha_G$ is less or larger than one.  For definiteness, we shall assume 
$\alpha_G > 1$. 
The maximal value for the QCD electric four-form field $E_G$ then depends on $\epsilon$ as follows\footnote{Note that we are not concerned with the actual value the gravitational Chern-Pontryagin density takes in the vacuum as it is not constrained by measurements.}
\be\label{eq:EQCDepsilon}
E= \Lambda^4\frac{\epsilon\alpha_G^2}{\epsilon \alpha_G^2 +\frac{\Lambda^4}{\Lambda_G^4}}\;.
\ee

The limit of massless neutrinos, $\epsilon\to 0$ leads to $E=E_G=0$. Thus, in this limit we recover the result of section \ref{sec:neutprot}.  In other words, massless neutrino fully protects the axion from gravity. 

Considering the value of the neutrino mass $m_\nu$ much smaller than the gravitational scale $\Lambda_G$, $m_\nu\ll \Lambda_G$, we get  $\epsilon\simeq\frac{m_\nu}{\Lambda_G}$. Then, the value of  $E$ in the QCD vacuum is instead
\be\label{eq:QCDthetawithsmallneutmasses}
E=\Lambda^4 \frac{\alpha_G^2m_\nu \Lambda_G^3}{\Lambda^4+\alpha_G^2m_\nu \Lambda_G^3}\;.
\ee
 In terms of $\theta$ this gives 
\be\label{eq:QCDthetawithsmallneutmasses}
\theta=\frac{\alpha_G^2m_\nu \Lambda_G^3}{\Lambda^4+\alpha_G^2m_\nu \Lambda_G^3}\;.
\ee
In order to be compatible with observations (cf. the electric dipole moment \cite{Shifman:1979if,Baker:2006ts}), $\theta$ must satisfy the bound $\theta<10^{-9}$. 

It is instructive to look at the values of $E$ for different choices of parameters. If the denominator in \eqref{eq:QCDthetawithsmallneutmasses} is dominated by $\alpha_G^2 m_\nu \Lambda_G^3$, there is essentially no screening of the four-form electric field, $E\sim  \Lambda^4$, and 
correspondingly the un-fine-tuned value is $\theta \sim 1$.  In this case, the axion solution of strong-CP problem is ruined. 
On the other hand, if $\Lambda^4 \gg \alpha_G^2 m_\nu \Lambda_G^3$, then requirement of the protection of the successful  axion 
mechanism, translates into the following bound on the lightest neutrino mass, 
\be\label{eq:neutbound}
 m_\nu\lesssim 10^{-9}\frac{\Lambda^4}{\alpha_G^2\Lambda_G^3}\;.
\ee
In turn, the experimental measurement of the neutrino mass would introduce an upper bound on the non-perturbative gravitational scale of the anomaly $\Lambda_G$. 
 
 The experimental searches \cite{katrinexp, GERDA} currently focus on the mass range $0.2 \,\mathrm{eV} < m_\nu < 2 \,\mathrm{eV}$. 
 A detection of the lightest neutrino mass in this window  would give the bound $\sqrt[3]{\alpha_G^2} \Lambda_G\lesssim 0.2\, \mathrm{GeV}$.

\section{Conclusions}
 In this work we have put forward yet another example of the highly profound connection between particle physics 
and non-perturbative quantum gravity.  The main players in this connection are the axion and the neutrino.   

Quantum gravity is believed to violate global symmetries, and among other things, 
ruin the axion solution of the strong-CP problem.  Usually, this impact is parameterized by 
introducing  all possible higher-dimensional operators suppressed by the Planck scale in the effective Peccei-Quinn Lagrangian \cite{russell}.  In such a picture, the impact is devastating. 
  However,  a closer look at the axion solution of the strong-CP problem reveals that it can be understood 
  as a gauge-Higgs effect in the QCD Chern-Simons three-form language.  In this formulation of the theory, because of the power of 
  gauge redundancy and the anomaly, it is possible to uniquely single out and fully parameterize the potentially dangerous 
  gravitational physics \cite{3,5}.  As we have seen, such physics can only come in form of a gravitational Chern-Simons 
  correlator,  or equivalently an effective gravitational three-form field that could  eat-up the axion. 
  
   By identifying the source of the danger, we were able to see the possible protection mechanism against it.
   This mechanism is built-in in the Standard Model in form of light neutrinos. 
    What we have shown is that, due to lepton number anomaly, the neutrino can sacrifice itself instead of the axion to a gravitational  three-form and neutralize its impact on the solution of the strong-CP problem. 
    This mechanism gives a phenomenological bound on the neutrino mass.  The precise measurement of this mass would reveal a bound on non-perturbative gravity scale. 
    
       Ideas displayed in this paper can be applied to other global approximate symmetries of the standard model, such as combinations of baryon and lepton numbers  along the lines of \cite{3}. 
  In particular, would be interesting to explore the consequences  of the electroweak analog of the 
  $\theta$-term, which must become physical after neutrino lepton number is broken by non-perturbative gravity  and $\eta_{\nu}$ becomes massive.                  
  
      Finally,  one of the consequences of the neutrino protection scenario is the existence of a new 
     effective low energy  pseudo-scalar degree of freedom, $\eta_{\nu}$,  that plays the role analogous to  $\eta'$ meson of QCD.   It would be interesting to explore its possible phenomenological  and cosmological consequences.

\section*{Acknowledgements}
It is pleasure to thank Cesar Gomez on valuable discussions on various aspects of axion physics. 
The work of G.D. was supported in part by Humboldt Foundation under Alexander von Humboldt Professorship, by European Commission under the ERC advanced grant 226371, by TRR 33 “The Dark Universe” and by the NSF grant PHY-0758032. The work of S.F. was supported by the Humboldt Foundation. The work of A.F. was supported by the FCT through the grant SFRH / BD / 77473 / 2011.

\end{document}